\documentclass[preprint,showpacs,preprintnumbers,nofootinbib,amsmath,amssymb]{revtex4}
\usepackage{graphicx}
\usepackage{dcolumn}
\usepackage{bm}
\begin{document}
\title{Shear-Viscosity to Entropy-Density Ratio\\ from Giant 
Dipole Resonances in Hot Nuclei}
\author{N. Dinh Dang$^{1,2}$}
  \email{dang@riken.jp} 
 \affiliation{1) Theoretical Nuclear Physics Laboratory, RIKEN Nishina Center
for Accelerator-Based Science,
2-1 Hirosawa, Wako City, 351-0198 Saitama, Japan\\
2) Institute for Nuclear Science and Technique, Hanoi, Vietnam}
\begin{abstract}
The Green-Kubo relation and fluctuation-dissipation theorem are 
employed to calculate the shear viscosity $\eta$ of a finite hot nucleus directly from 
the width and energy of the giant 
dipole resonance (GDR) of this nucleus. The 
ratio $\eta/s$ of shear viscosity $\eta$ to entropy density $s$ is
extracted from the experimental systematics of the GDR
in copper, tin and lead isotopes at finite temperature $T$.  
These empirical results are then compared with the  
predictions by several independent models, as well as with almost 
model-independent estimations. Based on these results, 
it is concluded that the ratio $\eta/s$ in medium and heavy 
nuclei decreases with increasing temperature $T$ to reach  
$(1.3 - 4)\times\hbar/(4\pi k_{B})$ at $T=$ 5 MeV.
\end{abstract}
\pacs{24.10.Pa, 24.30.Cz, 24.60.Ky, 25.70.Gh, 25.75.Nq}
\keywords{Suggested keywords}
\maketitle
\section{Introduction}
The recent observations of the charged particle elliptic flow and jet 
quenching in 
ultrarelativistic Au-Au and Pb-Pb collisions performed at 
the Relativistic Heavy Ion Collider
(RHIC)~\cite{RHIC} at Brookhaven National Laboratory 
and Large Hadron Collider (LHC)~\cite{LHC} 
at CERN have been the key experimental discoveries in the creation and 
study of quark-gluon plasma (QGP). The analysis of the 
data obtained from the hot and dense system produced in these 
experiments revealed that the strongly interacting matter formed in these collisions  
is a nearly perfect fluid with extremely low viscosity. 
In the verification of the condition for applying hydrodynamics to 
nuclear system, it turned out that the quantum mechanical 
uncertainty principle requires a finite viscosity for any thermal 
fluid. In this respect, one of the most fascinating theoretical 
findings has been the conjecture by Kovtun, Son and Starinets (KSS)~\cite{KSS} 
that the ratio $\eta/s$ of shear viscosity $\eta$ to the entropy 
volume density $s$ is bounded below for all fluids, namely the value
\begin{equation}
    \frac{\eta}{s}= \frac{\hbar}{4\pi k_{B}}\simeq 5.24\times 
    10^{-23}\hspace{1mm}  
    {\rm Mev}\hspace{1mm}  {\rm s}~.
    \label{KSSbound}
    \end{equation}
    is the universal lower bound (the so-called KSS bound or KSS unit).
    Although several theoretical counter examples have been proposed, 
no fluid that violates this lower bound 
has ever been experimentally found so far~\cite{Schafer}. 
The QGP fluid produced at RHIC has $\eta/s\simeq$ (2 - 3) KSS units.
Given this conjectured universality, there has been an increasing 
interest in calculating the ratio $\eta/s$ in different systems.

The first theoretical study that calculated the ratio $\eta/s$ for finite nuclei
has been the recent article by Auerbach and Shlomo, who estimated 
$\eta/s=$ (4 - 19) and (2.5 - 12.5) KSS units for heavy and light nuclei,
respectively~\cite{Auerbach}. These results have been obtained within 
the framework of the Fermi liquid drop model (FLDM)~\cite{FLDM}, 
which was applied to the damping of giant collective vibrations. 
The calculated shear viscosity $\eta$ in this study increases with temperature $T$ up to
quite high value of $T\sim$ 10 MeV to reach a
maximum at $T\sim$ 12 - 13 MeV. At higher $T$ a decrease of $\eta$ is 
seen. As a result, within the region
0 $\leq T\leq$ 5 MeV, where giant resonances exists, the 
damping width predicted by the FLDM always increases with $T$, 
being roughly proportional to $\eta$~\cite{FLDM,Auerbach1}.
Such temperature dependence contradicts the experimental systematics 
of the width of the giant dipole resonance (GDR) in hot nuclei. 
As a matter of fact, a large number of experiments on heavy-ion 
fusion and inelastic
scattering of light projectiles on heavy targets has shown that, 
while the location of the GDR peak (the GDR energy) is rather 
insensitive to the variation of $T$, its full width at the half 
maximum (FWHM) increases with $T$ only within 1 $\leq T\leq$ 2.5 MeV.
Below $T\sim$ 1 MeV, it remains nearly constant, whereas 
at $T>$ 3 - 4 MeV the width seems to saturate~\cite{Bracco,
Enders, Baumann, Heckmann, Kelly,Pb200}. 
To calculate the ratio $\eta/s$, the authors of Ref. \cite{Auerbach} employed 
the Fermi gas formula for the entropy $S=2aT$ with a
temperature-independent level density parameter $a$. This
approximation too is rather poor for finite nuclei, as has been
pointed out by one of the authors of Ref. \cite{Auerbach}, who has 
proposed a fitting formula for the temperature-dependent density 
parameter $a(T)$~\cite{Shlomo}.  Therefore, although 
the ratio $\eta/s$, which was obtained in Ref. \cite{Auerbach} by
dividing two increasing with $T$ quantities, does decrease
qualitatively to reach a values within one order of the KSS bound as 
$T$ increases up to 2 - 3 MeV, it is highly desirable to obtain a 
refined quantitative estimation for this ratio in finite hot nuclei 
from both theoretical and experimental points of view. 

The aim of the present work is to calculate the ratio
$\eta/s$ directly from the most recent and accurate 
experimental systematics of the GDR widths in hot nuclei. 
The extracted empirical values are then 
confronted with theoretical predictions by four models, which have 
been developed to describe the temperature dependence of the GDR width, namely the phonon damping
model (PDM)~\cite{PDM1,PDM3,PDM2}, two thermal shape fluctuation 
models (TSFM)~\cite{Ormand,Kusnezov}, as well as the FLDM mentioned above. 
An attempt to pinpoint the high-temperature limit of the ratio $\eta/s$ 
in finite nuclei in the most 
model-independent way is also undertaken.

The article is organized as follows. The formalism of calculating the 
shear viscosity $\eta$ from the GDR width and energy is discussed in 
Sec. \ref{visco}. The theoretical assessment for the entropy density 
is given in Sec. \ref{entro}. The analysis of numerical results is presented 
in Sec. \ref{results}. The article is summarized in the last section, 
where conclusions are drawn.
\section{Shear viscosity}
\label{visco}
\subsection{Shear viscosity at zero temperature}
\label{theor}
The increase of widths of nuclear giant resonances with decreasing the mass 
number suggests that the damping mechanism of collective vibrations 
might qualitatively be similar to that of a viscous fluid, where 
damping of sound waves under the influence of viscosity 
increases as the system volume decreases~\cite{Lamb}. 
From the viewpoint of collective theories, one of the fundamental 
explanations for 
the giant resonance damping remains the friction term (i.e. viscosity) 
of the neutron and proton fluids~\cite{Auerbach1}. 
A quantitative description of the 
dissipative behavior requires an inter-particle collision term to be 
included into the equation of motion for the one-body density matrix. 
For example, the nuclear fluid-dynamics approach incorporated a collision term in 
the Landau-Vlasov equation to derive 
the momentum conservation, which includes three terms similar to
the stress tensor, shear modulus, and dissipative component of the 
momentum flux tensor, respectively~\cite{FLDM}. The latter resembles the viscous 
term in the macroscopic Navier - Stokes equation, and 
is proportional to the damping coefficient of 
collective motion in the regime of rare collisions (zero-sound regime). 
Viscosity has also been employed to describe the decay of collective excitations 
in the context of nuclear fission in the 1970s~\cite{Swia}. 

In the microscopic description, 
the (quantal) width $\Gamma_{Q}(0)$ of giant resonances at $T=$ 0 
($\sim$ 4 - 5 MeV in medium and heavy nuclei) consists of 
the Landau width $\Gamma^{LD}$, spreading width $\Gamma^{\downarrow}$, 
and escape width $\Gamma^{\uparrow}$. The Landau width $\Gamma^{LD}$ is essentially 
the variance $\sigma = \sqrt{\langle E^{2}\rangle - \langle 
E\rangle^{2}}$ of the distribution of 
particle-hole ($ph$) states forming the giant 
resonance. The spreading width $\Gamma^{\downarrow}$ is caused by coupling of $1p1h$ states 
to more complicate configurations, first of all, the $2p2h$ ones, 
whereas the escape width $\Gamma^{\uparrow}$ arises because of 
coupling to the continuum causing the direct particle decay into hole 
states of the residual nucleus. In medium and heavy nuclei,
$\Gamma^{LD}$ and $\Gamma^{\uparrow}$ account only for a small fraction 
of the total width $\Gamma_{Q}(0)$. The major 
contribution  is given by 
$\Gamma^{\downarrow}$. In light nuclei, 
$\Gamma^{\uparrow}$ gives a dominant contribution, 
whereas $\Gamma^{LD}$ is also mainly apparent. 
Within the semiclassical approaches such as the 
Landau-Vlasov kinetic theory~\cite{alpha} or phenomenological approaches to 
nuclear friction~\cite{Fiolhais}, $\Gamma^{LD}$ corresponds 
to the collisionless damping or one-body dissipation (long-mean free 
path), whereas $\Gamma^{\downarrow}$ comes from the collision damping 
or two-body dissipation (short-mean free path). In the hydrodynamic 
theory of collective motion, which is based on a short-mean free path, 
the dissipative effects are usually bulk phenomena 
caused by the viscous shearing stresses between adjacent layers of fluid. 
The microscopic mechanism of this energy 
dissipation resides in the coupling of $1p1h$ configurations to $2p2h$ 
ones, which causes the spreading width $\Gamma^{\downarrow}$ of giant resonances.
This is how the shear viscosity is related to the damping of collective 
motion due to two-body interaction between nucleons in nuclei or 
molecules of a fluid. 

The one-body dissipation (long-mean free path) 
has been introduced based on the argument that 
the nucleon mean-free path is long compared to 
the nuclear radius. It arises primarily from the 
collisions of nucleons with the moving nuclear surface rather 
than with each other (the wall formula)~\cite{wall}. Although neither  
the wall formula nor the ordinary 
two-body viscosity can describe correctly the experimental widths of 
giant resonances, 
the predictions by the two-body viscosity (short-mean free path) 
are much closer to the experimental data~\cite{Nix}.
As for the fission-fragment kinetic energies, the results obtained on 
the basis of one-body dissipation agree with the experimental values 
equally well as those predicted by two-body 
viscosity~\cite{fission}. 
The evidence shows that 
a comprehensive view of the damping of giant resonances is likely a 
sum of one- and two-body contributions. This is consistent with the 
microscopic picture, where the one-body dissipation is described within the 
random-phase approximation (RPA), whereas the two-body dissipation is 
taken into account by coupling of the $1p1h$ states obtained within 
the RPA to $2p2h$ configurations or collective phonon beyond the RPA.

The discussion above indicates an uncertainty 
in extracting the value $\eta(0)$ of the shear viscosity $\eta(T)$ at 
$T=$ 0, given different dissipation mechanisms. In 
Ref.~\cite{Auerbach1}, the two-body viscosity was employed 
under the assumption of a rigid nuclear boundary to fit the data of 
isovector and isoscalar giant resonances at $T=$ 0. A value 
$\eta(0)\simeq$ 1$u$ $\simeq$ 0.016 TP (terapoise) has been found, 
where $u = 10^{-23}$ Mev s fm$^{-3}$. 
The analysis of nuclear fission data based on the two-body collisions
~\cite{fission} gives $\eta(0)$ 
in the range of (0.6 - 1.2)$u$, or (0.01 - 0.02) TP, under the assumption that 
scission occurs at zero radius of the neck rupture. 
A later study~\cite{fission1} assumed a finite radius for the neck 
rupture and found larger $\eta(0) =$ (1.1 - 2.5)$u$, or (0.02 - 0.04) TP. 
The authors of Ref.~\cite{Nix} adopted $\eta(0)\simeq$ (1.9 $\pm$ 
0.6)$u$, or 0.03 
$\pm$ 0.01 TP, to calculate 
the widths of giant resonances for nuclei with deformable surfaces under 
the assumption of incompressible, irrotational, small-amplitude 
nuclear flow. The predicted theoretical widths are 3 times larger than the experimental 
values within the one-body disspation mechanism based on the wall formula. 
For a modified one-body disspation, the calculated widths are 
smaller than the experimental ones. In Ref. \cite{Nix1}, the same authors 
calculated the fission-fragments kinetic energies using two-body 
viscosity in a similar way as that of Ref. \cite{Nix}, but with a 
modified potential. They found that the value $\eta(0)=$ 0.936$u$ (0.015 TP) 
satisfactorily reproduces the experimental data. This value is very 
close to that obtained in Ref. ~\cite{Auerbach1}. 
The authors of Ref. \cite{Vanin} pointed out that anomaly 
large values of $\eta(0)$, in the range of (2 - 25)$u$, must be used in order to 
obtain a simultaneous description of the variances of mass 
distributions and multiplicities of prescission particles on the 
basis of both one- and two-body dissipations. The 
strong disagreement between the largest value 
$\eta(0)=$ 25$u$ obtained in this case and thoses given in other 
references mentioned above shed doubt on the posibility of consistently 
describing the mass-energy distribution and prescission-particle 
multiplicity.  

In the present article, the value $\eta(0)=$ 1$u$, extracted in
Ref.~\cite{Auerbach1}, is adopted as a parameter in combination with the lower and upper bounds, 
equal to 0.6$u$ and 1.2$u$, respectively, obtained in Ref. 
\cite{fission} and applied here as error bars. 
The justification of this choice is based on two reasons. The first one is that the present 
article considers the evolution of $\eta(T)$ as a function of $T$ 
based on the GDR in hot heavy (spherical or weaky deformed) nuclei. At 
$T=$ 0 it should naturally be equal to $\eta(0)$ extracted from 
fitting the ground-state GDR (i.e. $T=$ 0) in Ref. \cite{Auerbach1}. 
Moreover, according to Ref. \cite{Fr}, compact 
nuclei favor the two-body viscosity, whereas the onset of one-body
dissipation is seen only in fissioning nuclei when the necking in 
starts, leading to a strong increase of the friction coefficient. This 
is also in line with the previously mentioned small constribution 
of $\Gamma^{LD}$ and $\Gamma^{\uparrow}$ in heavy nuclei. The second reason is that the 
present article attempts to see how low the ratio $\eta/s$ can go 
with increasing $T$, or how the KSS limit is fulfilled in hot nuclei. 
The lowest value of $\eta(0)$ found in the above-mentioned estimations is $\eta(0)=$ 
0.6$u$~\cite{fission}. The same lower bound  
has also been adopted in Ref. \cite{Auerbach} to calculate $\eta(T)$ 
within the FLDM, where the upper bound varies within the range 
of (1.9 $\pm$ 0.6)$u$, i.e. the same as used in Ref.~\cite{Nix}.
As has been mentioned above, these large upper bounds fail to 
reproduce the giant resonance width.
\subsection{Theoretical description of temperature dependence of shear viscosity $\eta(T)$}
\label{theor2}
With these cautions regarding the selected values $\eta(0)$, 
one now proceeds to study the evolution of $\eta(T)$ as a 
function of $T$.  The energy dissipation, 
which is a characteristic of a non-equilibrium or local thermodynamic 
equilibrium state (such as electrical conductivity, heat 
diffusion, shear viscosity,\ldots) is related to fluctuations in 
statistical equilibrium or global thermodynamics equilibrium  (such as thermal noise of 
electric and heat currents, collective 
vibrations,\ldots) by means of the 
fluctuation-dissipation theorem (FDT)~\cite{Kubo2,Zubarev}. 
This is realized making use of 
the Green-Kubo formula~\cite{Kubo1}, which is an exact expression for the linear 
transport coefficient of any system at a given temperature $T$ and 
density $\rho$ in terms of the time dependence of equilibrium fluctuations in the conjugate flux.
The Green-Kubo formula expresses
the shear viscosity $\eta(T)$ in terms of the correlation function of
the shear stress tensors $T_{xy}(t,{\bf x})$ as
\begin{equation}
    \eta(T) = \lim_{\omega\rightarrow 0}\frac{1}{2\omega}\int dt d{\bf
    x}e^{i\omega t}\langle[T_{xy}(t,{\bf x}),T_{xy}(0,0)]\rangle~,
    \label{Kubo}
    \end{equation}
    where the average $\langle\ldots\rangle$ is carried out within 
    a equilibrium statistical ensemble, such as the grand canonical 
    ensemble in the present article.
    From the FDT, it follows that the integrated 
    expression divided by $2\omega$ at  
    the right-hand side of Eq. (\ref{Kubo}) is proportional to the absorption 
cross section $\sigma(\omega,T)$. Therefore the following identity 
holds 
\begin{equation}
   \eta(T) = \lim_{\omega\to 0}\frac{1}{2\omega 
   i}[G_{A}(\omega)-G_{R}(\omega)]=-\lim_{\omega\to 0}\frac{{\rm Im} G_{R}(\omega)}{\omega}
   =\lim_{\omega\to 0}\frac{\sigma(\omega,T)}{C}~,
    \label{Nyquist}
    \end{equation}        
    where $G_{A}(\omega)$ and $G_{R}(\omega)$ are the advanced and 
    retarded Green functions, respectively, with $G_{R}(\omega) = 
    -i\int dt d{\bf
    x}e^{i\omega t}\theta(t)\langle[T_{xy}(t,{\bf x}),T_{xy}(0,0)]\rangle$, 
    and $G_{A}(\omega) = G_{R}(\omega)^{*}$. This relation has been employed in 
the anti-de Sitter/Conformal field theory to derive the KSS 
conjecture~\cite{KSS}, where $C$ is equal to $16\pi G$ with 
    $G$ being the ten-dimensional gravitational constant. 
    It has been shown in Ref. \cite{KSS,Das} that the graviton absorption cross section 
    $\sigma(\omega)$, used at the right-hand side of Eq. 
    (\ref{Nyquist}), must not vanish in the 
    zero-frequency limit ($\omega\to 0$) for nonextremal black 
    branes, and is actually equal to the area of horizon, so 
    that one can use $\sigma(0)$ to obtain the shear viscosity of the hot 
    supersymmetric Yang-Mills plasma. That Eqs. (\ref{Kubo}) 
    and (\ref{Nyquist}) indeed contain one-body dissipation has been 
    shown, e.g., by the authors of Ref. \cite{JPA}, who 
    derived the wall formula~\cite{wall} as the small-frequency 
    limit of the FDT.
    
In finite nuclei, the GDR photoabsorption cross
section is quantum mechanically 
described by the Breit-Wigner distribution from the Breit-Wigner's 
theory of damping~\cite{BW}
\begin{equation}
    \sigma_{GDR}(\omega) =\sigma_{GDR}^{int}f^{\rm 
    BW}(\omega,E_{GDR},\Gamma)~,\hspace{2mm} 
  f^{\rm BW}(\omega,E_{GDR},\Gamma)=\frac{1}{\pi}
  \frac{\Gamma/2}{[(\omega-E_{GDR})^{2}+(\Gamma/2)^{2}]}
\label{sigmaGDR}
\end{equation}
where $\sigma_{GDR}^{int}=
(1+k)\times TRK$ is the GDR integrated cross section 
with the Thomas-Reiche-Kuhn sum rule $TRK = 60 NZ/A$
(MeV mb), $\Gamma$ is FWHM of the GDR, and $E_{GDR}$ is its energy. 
The enhancement factor $k\simeq$ 0.5 - 0.7 
represents the additional strength, $k\times TRK$, 
above the GDR and below the meson threshold at $\sim$ 
140 MeV, which is usually attributed to the contribution due to 
meson-exchange forces.
Defining $C$ as a normalization factor to 
reproduce the value $\eta(0)$ at $T=$ 0 as
\begin{equation}
     C = \frac{\lim_{\omega\to 
     0}[\sigma_{GDR}(\omega,T=0)]}{\eta(0)}~,
     \label{C}
     \end{equation} 
     and inserting it as well as the 
right-hand side of Eq. 
(\ref{sigmaGDR}) into that of 
(\ref{Nyquist}), one obtains the final expression for the 
shear viscosity at temperature $T$ in the form
\begin{equation}
\eta(T)=\eta(0)\frac{\Gamma(T)}{\Gamma(0)}
\frac{E_{GDR}(0)^{2}+[\Gamma(0)/2]^{2}}{E_{GDR}(T)^{2}+[\Gamma(T)/2]^{2}}~.
\label{eta1}
\end{equation}
In principle, Eq. (\ref{eta1}) is not limited to the GDR, but can 
also be applied to calculate the temperature dependence of the 
transport coefficient in any transport process if 
its resonance scattering cross section is known. It is clear from Eq. 
(\ref{eta1}) that, unlike the prediction by nuclear hydrodynamic 
theories (e.g. Ref. ~\cite{FLDM}), $\eta(T)$ is not proportional to the 
GDR width $\Gamma(T)$, but is an infinite geometric series of 
$x(T)\equiv\Gamma(T)/[2E_{GDR}(T)]$, namely $\eta(T) 
=\eta(0)[x(T)/x(0)][1+x^{2}(0)]\sum_{n=0}^\infty(-)^{n}x^{2n}(T)$ [$x(T)<$ 1]. 
It is proportional to $\Gamma(T)$ only in the limit of 
small damping ($x\ll$ 1, the hydrodynamic regime), when Eq. (\ref{eta1}) 
reduces to
\begin{equation}
\eta(T)\simeq\eta(0)\frac{\Gamma(T)}{\Gamma(0)}~,
\label{small-damp}
\end{equation}
under the assumption that $E_{GDR}$ does not depend on $T$. 
This limit can be independently verified using the Stokes 
law of sound attenuation $\alpha$, according to which 
$\alpha=2\eta\omega^{2}/(3\rho V^{3})$. Indeed, by 
using the relation $\alpha = 2\Gamma(T)/v$, one obtains $\Gamma(T) = 
\eta(T) v\omega^{2}/(3\rho V^{3})$. Knowing $\eta(0)$ and $\Gamma(0)$, 
one can determine $v=3\rho V^{3}\Gamma(0)/[\eta(0)\omega^2]$. 
Inserting this expression of $v$ into that of $\Gamma(T)$, one recovers
the limit (\ref{small-damp}).    

Because the GDR strength function in microscopic theories of the GDR 
damping is usually described with a single Breit-Wigner distribution 
or a superposition of them,
Eq. (\ref{eta1}) will be used in the present article to calculate the
shear viscosity within the PDM and TSFM. It is worth mentioning 
that definition (\ref{C}) avoids the necessity of requiring 
$\sigma(0,T)\neq$ 0, because even with $\sigma(0,T)=$ 0, inserting Eq. 
(\ref{C}) into the right-hand side of Eq. 
(\ref{Nyquist}) yields the $0/0$-type limit for $\eta(T)$ (at 
$\omega\to 0$), which can be finite. 
This is actually the case, which will be discussed later 
in Sec. \ref{exp}, when the Lorentz distribution 
is used instead of the Breit-Wigner one (\ref{sigmaGDR}) 
to fit the photoabsorption cross section.

Equation (\ref{eta1}) shows that, given the values
$\eta(0)$, GDR width $\Gamma(T)$ 
and energy $E_{GDR}(T)$ at zero and finite $T$, one can calculate 
the shear viscosity $\eta(T)$ as a function of $T$. Considering the evolution 
of the GDR width as a function of $T$ under the assumption 
that the microscopic mechanism of the 
quantal width of the GDR at $T=$ 0 is known, 
the present article adopts the predictions by four 
models, namely the PDM~\cite{PDM1,PDM3,PDM2}, two versions of TFSM, namely the adiabatic 
model (AM)~\cite{Ormand} and the phenomenological parametrization of 
the TFSM, referred to as pTSFM hereafter~\cite{Kusnezov}, and the 
FLDM~\cite{Auerbach,FLDM}. Because these models have already been 
discussed in great details in Refs. 
\cite{PDM1,PDM2,PDM3,Ormand,Auerbach,FLDM}, only their main features 
and/or results, used in  the present article, are summarized below.
\subsubsection{Phonon damping model}
The PDM employs a model Hamiltonian, which consists of the 
independent single-particle 
(quasiparticle) field, GDR phonon field, and the coupling between 
them [See Eq. (1) in Ref. \cite{PDM1}, e.g.]. The Woods-Saxon potentials for
spherical nuclei at $T=$ 0 are used to obtain the single particle energies.
These single-particle spectra span a large space from around $-40$ 
MeV up to around 17 - 20 MeV. They are kept unchanged with $T$ based 
on the results of the temperature-dependent selfconsistent 
Hartree-Fock calculations, which showed that the single-particle 
energies are not sensitive to the variation of $T$ up to $T\sim$ 6 - 
7 MeV in medium and heavy nuclei~\cite{Quentin}.  
The GDR width $\Gamma(T)$ is given as the sum of
the quantal width, $\Gamma_{\rm Q}$, and thermal width, $\Gamma_{\rm T}$:
\begin{equation}
\Gamma(T)=\Gamma_{\rm Q}+\Gamma_{\rm T}~.
\label{widthtotal}
\end{equation}
In the presence of superfluid pairing, the quantal and thermal widths 
are given as~\cite{PDM2}
\begin{equation}
\Gamma_{\rm Q}=2\pi F_{1}^{2}\sum_{ph}[u_{ph}^{(+)}]^{2}(1-n_{p}-n_{h})
\delta[E_{\rm GDR}(T)-E_{p}-E_{h}]~,
\label{GammaQ}
\end{equation}
\begin{equation}
\Gamma_{\rm T}=2\pi F_{2}^{2}\sum_{s>s'}[v_{ss'}^{(-)}]^{2}(n_{s'}-n_{s})
\delta[E_{\rm GDR}(T)-E_{s}+E_{s'}]~,
\label{GammaT}
\end{equation}
where $(ss')$ stands for $(pp')$ and $(hh')$ with $p$ and $h$ denoting the orbital 
angular momenta $j_{p}$ and $j_{h}$ for particles and holes, 
respectively. Functions $u_{ph}^{(+)}$ and $v_{ss'}^{(-)}$ are 
combinations of the Bogoliubov coefficients
$u_{j}$, $v_{j}$, namely $u_{ph}^{(+)}=u_{p}v_{h}+v_{p}u_{h}$, and
$v_{ss'}^{(-)}=u_{s}u_{s'}-v_{s}v_{s'}$. The quantal width 
is caused by coupling of the GDR vibration (phonon)
to noncollective $ph$ configurations with the factors
$(1-n_{p}-n_{h})$, whereas the thermal width arises due to coupling of 
the GDR phonon to $pp$ and $hh$ configurations including the factors
$(n_{s}-n_{s'})$ with $(s,s') = (h,h')$ or $(p,p')$. 
The quasiparticle occupation number $n_{j}$ has the shape of 
a Fermi-Dirac distribution
\begin{equation}
n_{j}^{FD}=[\exp(E_{j}/T)+1]^{-1},
\label{njFD}
\end{equation}
smoothed with a Breit-Wigner kernel, 
whose width is equal to the quasiparticle damping with the quasiparticle energy
$E_{j}=\sqrt{(\epsilon_{j}-\lambda)^{2}+\Delta(T)^{2}}$ [See Eq. (2) of
\cite{PDM2}]. Here 
$\epsilon_{j}$, $\lambda$, and $\Delta(T)$ are the (neutron or proton) 
single-particle energy, chemical potential, and pairing gap, respectively. When
 the quasiparticle damping is small, as usually the 
 case for GDR in medium and heavy nuclei, 
 the Breit-Wigner-like kernel can be replaced with 
 the $\delta$-function so that the quasiparticle occupation number  
 $n_{j}$ can be approximated with the Fermi-Dirac
 distribution $n_{j}\simeq n_{j}^{\rm FD}$ of non-interacting 
 quasiparticles. The PDM predicts 
 a slight decrease of the quantal width (in agreement with the 
 finding that the Landau and spreading widths of GDR 
 do not change much with $T$~\cite{spread}), and a strong increase of the thermal width 
with increasing $T$, as well as 
a saturation of the total width at $T\geq$ 4 - 5 MeV in tin and
lead isotopes~\cite{PDM1} in agreement with experimental 
systematics~\cite{Bracco, Enders, Baumann, Heckmann, Kelly,Pb200}. 

For the open-shell nuclei pairing parameters $G$ are chosen for neutrons and/or protons  
to reproduce the empirical values at $T=$ 0 for the 
neutron and/or proton pairing gaps $\Delta(0)$.  
In the presence of strong thermal fluctuations, 
the pairing gap $\Delta(T)$ of a finite nucleus does not collapse at the critical 
temperature $T_{c}$, corresponding to the superfluid-normal phase
transition predicted by the BCS theory for infinite systems, but
decreases monotonically as $T$ increases~\cite{Moretto,MBCS,SCQRPA,Dean}.
The effect due to thermal fluctuations of quasiparticle numbers, 
which smooths out the superfluid-normal phase transition, is taken 
into account by using $\Delta(T)$ 
obtained as the solution of the modified BCS (MBCS) equations ~\cite{MBCS}.
The use of the MBCS thermal pairing gap $\Delta(T)$ 
for $^{120}$Sn leads to a nearly constant GDR width or even a slightly 
decreasing one at $T\leq$ 1 MeV  ~\cite{PDM2} in
agreement with the data of Ref. \cite{Heckmann}. 

It is worth noticing that, within the PDM, the GDR strength function is calculated  
in terms of the GDR spectral intensity $J_{q}(\omega)\propto
-2{\rm Im}[G_{R}(\omega)]/[\exp(\omega/T)-1]$ with $G_{R}(\omega)$ 
being the retarded Green function associated with the GDR. Its final 
form reads
\begin{equation}
J_{q}(\omega)={f^{\rm BW}(\omega, 
\omega_{q}',2\gamma_{q})}[e^{\omega/T}-1]^{-1}~,
\label{Jq}
\end{equation}
with $\omega'_{q}=\omega_{q}+P_{q}(\omega)$, where $\omega_{q}$ is the unperturbed 
phonon energy, $P_{q}(\omega)$ is the polarization operator 
arised due to coupling of GDR phonon to $ph$, $pp$ 
and $hh$ configurations. The GDR energy is defined as the 
solution of the equation 
\begin{equation}
\omega - \omega_{q} - P_{q}(\omega) = 0,
\label{EGDR}
\end{equation}
at which one obtains $\Gamma(T)=2\gamma_{q}$ in Eq. 
(\ref{widthtotal}). The use of 
Eq. (\ref{Jq}) within the PDM then yields 
exactly Eq. (\ref{Nyquist}). 
The PDM as well as the selection of its parameters $F_{1}$ and $F_{2}$ 
in Eqs. (\ref{GammaQ}) and (\ref{GammaT}) are presented and discussed thoroughly in Refs. 
\cite{PDM1,PDM3,PDM2} and references therein, to which the reader is 
referred for further details.
\subsubsection{Adiabatic model (AM)}
The AM~\cite{Ormand} assumes that the time scale for thermal fluctuations is slow 
compared to the shift of the dipole frequency caused by the 
fluctuations so that the the GDR strength function can be averaged 
over all quadrupole shapes with deformation $\alpha_{2\mu}$ and orientations. The angular-momentum 
projected GDR cross section $\sigma(\omega)$ at a given temperature $T$ is calculated 
within the AM as a thermal average over the shape-dependent cross 
sections $\sigma(\omega,\alpha_{2\mu},{\omega}_{J})$
\begin{equation}
\sigma(\omega) = \frac{1}{Z_{J}}\int\frac{{\cal D}[\alpha]}{{\cal 
I}(\beta,\gamma,\theta,\psi)^{3/2}}
\sigma(\omega,\alpha_{2\mu},{\omega}_{J})\exp[-F(T,\alpha_{2\mu},\omega_{J})/T]~,
\label{AM}
\end{equation}
where $\omega$ is the photon energy, 
${\cal D}[\alpha]=\beta^{4}\sin(3\gamma)d\beta d\gamma d\Omega$ 
is the volume element, $Z_{J}=\int{\cal D}[\alpha]{\cal 
I}^{-3/2}\exp[-F(T,\alpha_{2\mu},\omega_{J})/T]$ 
is the partition function, ${\cal I}(\beta,\gamma,\theta,\psi)=
I_{1}\cos^{2}\psi\sin^{2}\theta + I_{2}sin^{2}\psi\sin^{2}\theta 
+I_{3}\cos^{2}\theta$ is the moment of inertia about the rotation axis, 
expressed in terms of the principal moments of inertia $I_{k}$ and 
the Euler angle $\Omega=(\psi,\theta,\phi)$, 
$F(T,\alpha_{2\mu},\omega_{J})=F(T,\alpha_{2\mu},0)+(J+1/2)^{2}/[2{\cal 
I}(\beta,\gamma,\theta,\psi)]$ is the free energy with 
$F(T,\alpha_{2\mu},0)$ denoting the cranking free energy 
at $\omega_{J}=$ 0. The free energy $F(T,\alpha_{2\mu},0)$ and the principal
moments of inertia are calculated using either the Nillson-Strutinsky
approach including shell corrections, or the liquid-drop model. 
The shape-dependent cross section 
$\sigma(\omega,\alpha_{2\mu},{\omega}_{J})$ is calculated at the 
saddle-point frequency $\omega_{J}= (J+1/2)/{\cal 
I}(\beta,\gamma,\theta,\psi)$, where the GDR is approximated as a 
rotating three-dimensional oscillator consisting of three fundamental 
modes with energies 
$E_{k}=70 A^{-1/3}\exp[-\sqrt{5/\pi}\beta\cos(\gamma+2\pi k/3)/2]$ 
($k = 1, 2, 3$). The GDR Hamiltonian in the intrinsic frame is written as $H_{GDR}=
\sum_{k}(p_{k}^{2}+E_{k}^{2}d_{k}^{2}) + 
\vec{\omega}_{rot}(\vec{d}\times\vec{p})$ where $d_{k}$ and $p_{k}$ 
are the coordinates and conjugate momenta of the GDR vibration, and 
$\vec{\omega}_{rot}$ is the rotation frequency. The GDR cross section 
in the intrinsic frame is calculated by using the Breit-Wigner 
distribution (\ref{sigmaGDR}) as
\[
    \sigma(\omega,\alpha_{2\mu},{\omega}_{J})=\sigma_{0}
    \sum_{\mu\nu}|\langle\nu|d_{\mu}|0\rangle|^{2}\omega \bigg[f^{\rm 
    BW}(\omega,E_{\nu},\Gamma_{\nu}) - f^{\rm 
    BW}(\omega,-E_{\nu},\Gamma_{\nu})\bigg]
    \]
    \begin{equation}
        =\sigma_{0}
    \sum_{\mu\nu}|\langle\nu|d_{\mu}|0\rangle|^{2}E_{\nu}f^{\rm 
    L}(\omega,E_{\nu}',\Gamma_{\nu})~,
    \label{sigmaAM}
    \end{equation}
where $\mu$ 
denote the spherical components of the dipole mode, $|\nu\rangle$ 
are the eigenstates of the model Hamiltonian, 
$\Gamma_{\nu}=\Gamma_{0}(E_{\nu}/E_{0})^{\delta}$ ($\nu = 1, 2, 3$) 
with $\delta=$ 1.8 are the parametrized intrinsic 
widths of the three components of the GDR, which are centered at $E_{\nu}$, 
whereas $E_{0}$ and 
$\Gamma_{0}$ are respectively the energy centroid and width of the 
GDR at $T=$ 0. Function $f^{\rm L}(\omega,E_{\nu}',\Gamma)$ 
is the Lorentz distribution
\begin{equation}
    f^{\rm L}(\omega,E_{\nu}',\Gamma) =
    \frac{\omega}{E_{\nu}}\bigg[f^{\rm BW}(\omega,E_{\nu},\Gamma) - f^{\rm 
    BW}(\omega,-E_{\nu},\Gamma)\bigg] 
    = \frac{2}{\pi}\frac{\omega^{2}\Gamma}
    {[\omega^{2}-(E_{\nu}')^{2}]^{2}+\omega^{2}\Gamma^{2}}~,
    \label{Lor}
    \end{equation}
    with $(E_{\nu}')^{2}=E_{\nu}^{2}+(\Gamma/2)^{2}$.
The normalization factor $\sigma_{0}$ ensures the integrated 
cross section of the GDR to be equal to the Thomas-Reich-Kuhn sum 
rule. 
The present article uses the GDR widths obtained within the 
AM for $^{120}$Sn and $^{208}$Pb as shown by solid lines in Figs. 1 
of Ref. \cite{Ormand}.
\subsubsection{Phenomenologically parametrized thermal shape fluctuation model (pTSFM)}
The pTSFM~\cite{Kusnezov} is essentially a phenomenological parametrization of the AM 
discussed in the previous section. This model proposes a 
phenomenological fit for the width of a liquid-drop GDR 
as a function of temperature 
$T$, mass number $A$ and angular momentum $J$. For $J\leq 20\hbar$ as
in the experimental systematics used in the present article, this 
phenomenological fit reduces to
\begin{equation}
    \Gamma(T,A) = \Gamma_{0}(A)+ c(A){\rm 
    ln}\bigg(1+\frac{T}{T_{0}}\bigg)~,
    \hspace{5mm} c(A) = 6.45 - A/100~.
    \label{pTSFM}
    \end{equation}
The reference 
temperature $T_{0}=$ 1 MeV is used in the pTSFM calculations. The 
present article uses Eq. (\ref{pTSFM}) to calculate the GDR width in 
copper, tin, and lead regions. The shell corrections within the 
Nillson-Strutinsky method are not included because they have almost 
no effect on the GDR width in open-shell nuclei, whereas for lead isotopes they are 
important only at $T\leq$ 1.2 MeV as shown in Fig. 4 of Ref. 
\cite{Kusnezov}.
\subsubsection{Fermi liquid drop model}
\label{FLDM}
The FLDM employs a collision kinetic equation, which includes the 
dissipative propagation of sound  wave in infinite nuclear matter, 
to directly calculate the shear viscosity $\eta$ 
as~\cite{Auerbach}
\begin{equation}
    \eta(T)=\frac{2}{5}\rho\epsilon_{F}\frac{\tau_{\rm 
    coll}}{1+(\omega\tau_{\rm coll})^{2}}~,\hspace{5mm} \tau_{\rm 
    coll}=\frac{\tau_{0}}{1+(\hbar\omega/2\pi T)^{2}}~,\hspace{5mm} 
    \tau_{0}=\hbar\alpha/T^{2}~.
    \label{etaFLDM}
    \end{equation}
   After inserting the explicit expressions for $\tau_{\rm coll}$ and 
    $\tau_{0}$, the expression for $\eta(T)$ becomes
\begin{equation}
    \eta(T)=\frac{2}{5}\rho\epsilon_{F}\frac{\hbar}{4\pi^{2}\alpha}
    \frac{1+(2\pi T/\hbar\omega)^{2}}{1+\{\hbar\omega[1+(2\pi 
    T/\hbar\omega)^{2}]/(4\pi^{2}\alpha)\}^{2}}~,
    \label{etaFLDM1}
    \end{equation}
The calculations within the FLDM used the 
Fermi energy $\epsilon_{F}=$ 40 MeV and the nuclear 
density $\rho = 0.16 fm^{-3}$~\cite{Auerbach}, whereas the parameter 
$\alpha$ has been estimated based on the 
    in-medium-nucleon-nucleon scattering cross section to be between
    around 9.2 for isoscalar modes and 4.6 for the isovector 
    ones~\cite{alpha}. The empirical giant resonance energy 
$\hbar\omega$ decreases from around 19 MeV to 13 MeV as the mass number 
$A$ increases from around 50 to 250~\cite{Berman}. Adopting these values, one 
    finds the factor $\{\hbar\omega/(4\pi^{2}\alpha)\}^{2}$ in the denominator of 
    the expression at the right-hand side of Eq. (\ref{etaFLDM1}) in 
    the range between 0.001 and 0.01. This shows that $\eta(T)$ can 
    be approximated at low $T$ with the zero sound 
    limit ($\omega\tau\gg$ 1, $T\ll\hbar\omega$) as~\cite{Auerbach,FLDM}
    \begin{equation}
    \eta(T)_{z.s}=\frac{2}{5}\rho\epsilon_{F}\frac{\hbar}{4\pi^{2}\alpha}
    \bigg[1+\bigg(\frac{2\pi T}{\hbar\omega}\bigg)^2\bigg]~.
    \label{zs}
    \end{equation}
    Using this limit, one can also readjust the parameter $\alpha$ to 
    reproduce the empirical values of $\eta(0)=$ 0.6$u$, 1.0$u$, 
    and 1.2$u$, discussed previously in Section 
    \ref{theor}. This leads to 
    $\alpha=$ 7.11, 4.27, and 3.56, respectively. 
    The lowest $\eta(0)\simeq$ 0.46$u$ was 
    obtained in Ref. \cite{Auerbach} by using $\alpha=$ 9.2.

The FLDM, however, offers only the expression for the 
collisional width, but not for the FWHM of the GDR at 
$T\neq$ 0 (See Eq. (333) of Ref. \cite{FLDM}) because it does not 
include the effect of collissionless damping (one-body dissipation). As a matter of fact, 
an attempt to fit this width with the total GDR 
width has resulted in a value of the cut-off factor $\bar{q}$, 
which is 4 times larger than the theoretically estimated 
realistic value $\bar{q}=$ 0.192~\cite{FLDM}. With Eq. (\ref{eta1}) 
proposed in the present article, one can readily derive the FWHM 
$\Gamma(T)$, knowing the values of other parameters, 
namely $r\equiv\eta(T)/\eta(0)$, $E_{GDR}(T)$, $E_{GDR}(0)$, and 
$\Gamma(0)$ by solving a simple quadratic equation for the unknown 
$\Gamma(T)$. As the result one obtains
\begin{equation}
    \Gamma(T) = \frac{4E_{GDR}(0)^{2} + 
    \Gamma(0)^{2}-\sqrt{[4E_{GDR}(0)^{2} 
    + \Gamma(0)^{2}]^{2}-[4rE_{GDR}(T)\Gamma(0)]^{2}}}{2r\Gamma(0)}~.
    \label{GFLDM}
    \end{equation}
The other solution (with the $+$ sign in front of the square root) is excluded because it does 
not give $\Gamma(T)=\Gamma(0)$ at $T=$ 0.  To have a real value of 
$\Gamma(T)$ by Eq. (\ref{GFLDM}), the expression under the square 
root at its right-hand side must not be negative. This leads to the 
constraint
\begin{equation}
    \frac{\eta(T)}{\eta(0)}\leq\frac{4E_{GDR}(0)^{2}+\Gamma(0)^{2}}{4E_{GDR}(T)\Gamma(0)}~.
    \label{ratio}
    \end{equation}
Based on the experimental systematics showing that $E_{GDR}(T)$ is not 
sensitive to the temperature change, one can put 
$E_{GDR}(T)\simeq E_{GDR}(0)$ in Eq. (\ref{ratio}). By using the fit 
$\Gamma(0)\simeq 0.3 E_{GDR}(0)$~\cite{Auerbach1}, it follows from Eq. (\ref{ratio}) 
that 
\begin{equation}
\frac{\eta(T)}{\eta(0)}\leq 3.41~.
\label{ratio2}
\end{equation}
This means that, while one can calculate the shear viscosity 
$\eta(T)$ from the width and energy of GDR from Eq. 
(\ref{eta1}) at any $T$, the inverse is not true, that is the GDR 
width $\Gamma(T)$ extracted from the same equation based on the 
values of shear viscosity $\eta(T)$ at zero and finite $T$ 
as well the values of $\Gamma(0)$ and $E_{GDR}(0)$ breaks down 
to become imaginary at a temperature $T_{c}$, starting from which 
$\eta(T)>3.41\eta(0)$. The width (\ref{GFLDM}) also 
depends on $\eta(0)$, hence, on the parameter $\alpha$, especially at 
high $T$ when $2\pi T\sim\hbar\omega$, as can be inferred from Eq. 
(\ref{etaFLDM1}).
\subsection{Empirical extraction of shear viscosity $\eta(T)$ at 
$T\neq$ 0}
\label{exp}
The experimental cross section of the GDR is 
often fitted with a Lorentz distribution $f^{\rm 
L}(\omega,E_{GDR},\Gamma)$ (\ref{Lor})~\cite{Berman}
    rather than
    with a Breit-Wigner one, $f^{\rm BW}(\omega,E_{GDR},\Gamma)$ 
    (\ref{sigmaGDR}). 
    The GDR energy in the Lorentz 
    distribution (\ref{Lor}) is defined as $E_{GDR}^{2} = 
    E_{D}^{2}+(\Gamma/2)^{2}$, where $E_{D}$ is the 
    energy of the dipole mode before switching on 
    the coupling to configurations that cause the GDR 
    width~\cite{Danos}. The nice fits obtained for a wide class 
    of GDRs built on the ground state ($T=$ 0) of 
    medium and heavy spherical nuclei seem to justify such {\it ad hoc} 
    practice. For the GDR at $T=$ 0, one has $E_{GDR}\gg$ 0 and 
$\Gamma\ll E_{GDR}$. 
Therefore the Breit-Wigner component centered at 
$-E_{D}$ at the right-hand side of Eq. (\ref{Lor}) for the 
Lorentz distribution has a negligible effect on the GDR shape, and can be safely 
neglected, leading to $f^{\rm L}(\omega,E_{GDR},\Gamma) \simeq
(\omega/E_{GDR})f^{\rm BW}(\omega,E_{GDR},\Gamma)$ 
    with $E_{GDR}\simeq E_{D}$. 
    Differences arise when $\Gamma$ becomes comparable with 
    $E_{GDR}$. Nonetheless the 
    Lorentz distribution has also been applied to fit the 
    experimentally measured GDRs in hot nuclei, where $\Gamma\sim 
    E_{GDR}$ at high $T$~\cite{Bracco,
Enders, Baumann, Heckmann, Kelly,Pb200}. 

Using the Lorentz distribution $f^{\rm L}(\omega,E_{\nu}',\Gamma)$, 
one has $\sigma(0,T)=$ 0 because of the multiplier 
$\omega^{2}$ at the right-hand side of Eq. (\ref{Lor}), which  
vanishes at $\omega\to$ 0. However,  the definition of the normalization factor 
    $C$ by Eq. (\ref{C}) guarantees the cancellation of this 
    multiplier $\omega^{2}$ in the expression for $\eta(T)$. Indeed,
    dividing $\sigma(\omega,T)$ by the normalization factor $C$ given 
    by Eq. (\ref{C}), and 
    then taking the limit $\omega\to$ 0, one obtains 
    the following expression for the shear viscosity
\begin{equation}
\eta^{\rm L}(T)=\eta(0)\frac{\Gamma(T)}{\Gamma(0)}
\bigg\{\frac{E_{GDR}(0)^{2}}{E_{GDR}(0)^{2}-[\Gamma(0)/2]^{2}+[\Gamma(T)/2]^{2}}\bigg\}^{2}~.
\label{etaLor}
\end{equation}
In the present article, in order to have an exhaustive comparison, both of Eqs. (\ref{eta1}) and 
(\ref{etaLor}) will be used to extract 
the empirical shear 
viscosity from the experimental systematics of the GDR widths and 
energies obtained in hot nuclei.
\section{Entropy density}
\label{entro}
The entropy density (entropy per volume $V$) is calculated as 
\begin{equation}
s = \frac{S}{V} = \rho\frac{S}{A}
\label{dens}
\end{equation}
with the nuclear density $\rho=$ 0.16 fm$^{-3}$. The entropy $S$ at 
temperature $T$ is 
calculated by integrating the Clausius definition of entropy as
\begin{equation}
    S = \int_{0}^{T}\frac{1}{\tau}\frac{\partial{\cal E}}{\partial\tau}d\tau~,
    \label{Clausius}
    \end{equation}
where ${\cal E}$ is the total energy of the system at temperature 
$\tau$, which is evaluated microscopically as within the PDM or macroscopically by 
using the Fermi gas formula, ${\cal E} = {\cal E}_{0} + aT^{2}$, as within the FLDM.

By taking the thermal average of the PDM Hamiltonian and applying 
Eq. (\ref{Clausius}), it follows that 
\begin{equation}
S=S_{F}+S_{B}~,
\label{Stotal}
\end{equation}
where $S_{F}$ and $S_{B}$ are the
entropies of the quasiparticle and phonon fields, 
respectively [See Eq. (1) of Ref. \cite{PDM1}].
The entropy $S_{\alpha}$ ($\alpha = F, B$) is given in units of Boltzmann
constant $k_{B}$ as
\begin{equation}
    S_{\alpha}^{\rm PDM} = -\sum_{j}N_{j}[p_{j}\ln
    p_{j}\pm (1\mp p_{j})\ln(1\mp p_{j})]~,
    \label{SPDM}
    \end{equation}
where $p_{j}=n_{j}$ are the quasiparticle 
occupation numbers ($\alpha = F$) or phonon occupation numbers
$p_{j}=\nu_{j}$ ($\alpha = B$), the upper (lower) sign is for quasiparticles
(phonons), $N_{j}= 2j+1$ and 1 for $\alpha = F$ and $B$, respectively.
For $\alpha = F$, the index $j$ denotes the single-particle 
energy level, corresponding to the orbital angular momentum $j$, whereas 
for $\alpha = B$, it corresponds to that of GDR phonon. 
Because the quasiparticle (single-particle) damping is negligible for 
heavy nuclei~\cite{PDM3}, 
it is neglected in the present calculations of entropy $S_{F}$ 
for the sake of simplicity, assuming $n_{j} = n_{j}^{FD}$ from Eq. 
(\ref{njFD}). Regarding the phonon occupation number for the GDR, it is 
approximated with the Bose-Einsten distribution $\nu_{GDR}\simeq\nu_{GDR}^{B}=
[\exp(E_{GDR}/T)-1]^{-1}$ in the present calculations.
This gives the upper bound for the entropy, hence the lowest bound 
for the ratio $\eta/s$, estimated within the PDM. Indeed, the phonon occupation number 
$\nu_{q}$ including the phonon damping is given by Eq. (2.34) 
of Ref. \cite{PDM3}. For the GDR ($q = GDR$) it is the Bose-Einstein distribution 
$\nu_{GDR}^{B}=[\exp(E_{GDR}/T)-1]^{-1}$ smoothed
with a Breit-Wigner kernel, whose width is equal to the GDR width, 
that is $\nu_{GDR}<\nu_{GDR}^{B}$.
Given $E_{GDR}\gg T$, it turns 
out, however, that $S_{B}\ll S_{F}$ so that in all the cases 
considered here, one has $S\simeq S_{F}$. For example, for $^{120}$Sn 
with $E_{GDR}\simeq$ 15.5 MeV and FWHM around 14 MeV at 
$T =$ 5 MeV~\cite{PDM1}, one finds $\nu_{GDR}^{B}\simeq 0.009$, which 
gives a negligible values 0.051 for $S_{B}$ as compared to
$S_{F}\simeq$ 109 (in units of $k_{B}$).

The Fermi gas formula for the entropy is
\begin{equation}
    S^{\rm FG} = 2aT~.
    \label{SFG}
    \end{equation}
This formula is used in the FLDM and the analysis of experimental 
data. The level density parameter $a=A/K$ with $K$ varying from 8 
and 13 - 14 when $A$ goes from the mass region of heavy nuclei 
to that of light ones. At $T\neq$ 0, the level density also depends on 
$T$~\cite{Shlomo,Ormand,Reis,Hagen}. The experimental temperature, 
width and energy of the GDR  were deduced by using the
temperature-dependent parametrization of the 
density parameter $a(T)$ shown as the dashed lines in the left and 
right panels of Fig. 4
of Ref. \cite{Baumann} for $^{120}$Sn and $^{208}$Pb, respectively.
The same parametrization will be used here to calculate the empirical entropy 
density $s$ from Eqs. (\ref{dens}) and (\ref{SFG}) for tin and lead 
isotopes. For $^{63}$Cu the empirically adopted value $A/a = 8.8$ MeV is used 
for the temperature range, where the GDR width was 
extracted~\cite{Cu59,Cu63}.
As for $S$ used within the FLDM, the value of the level density 
parameter $a$ that fits best the microscopic and empirical entropies 
will be adopted in the calculations. Regarding the entropy 
used for calculating $\eta/s$ within the AM and pTSFM, although the precise one 
should be obtained from Eq. (\ref{Clausius}), the same 
entropy (density) as that used for the FLDM will be adopted because only the 
liquid-drop version of these models is considered here.
\section{Analysis of numerical results}
\label{results}
\subsection{GDR width}
\label{widthresults}
Within the PDM the GDR width obtained for $^{120}$Sn including the 
effect of thermal pairing in Ref.\cite{PDM2} is employed [the thick 
dotted line in Fig. 4 (a) of Ref. \cite{PDM2}], whereas for 
$^{208}$Pb the results of Ref. \cite{PDM1} are used [the solid line 
with diamonds in Fig. 1 (b) of Ref. \cite{PDM1}].
For $^{63}$Cu, the effect of pairing on the GDR width is small so it 
is not included in the GDR width calculations, which 
are carried out here for the first time within the PDM.
The values $F_{1}=$ 0.332 MeV and
$F_{2}=$ 0.933 MeV are chosen for this nucleus to reproduce a stable 
$E_{GDR}\simeq$ 16 - 17 MeV
as $T$ varies up to 5 MeV, and the FWHM equal to around 7 MeV at $T<$ 
0.5 MeV in agreement with the experimental values of the 
grounds-state GDR. The GDR widths predicted within the AM for 
$^{120}$Sn and $^{208}$Pb 
are read from the solid and dotted lines of Fig. 5 of 
Ref.~\cite{Ormand}, respectively, because only the liquid-drop 
version of this model is considered here (For $^{120}$Sn shell 
corrections have a negligible effect on the GDR width as shown in Fig. 
4 of Ref. \cite{Kusnezov}). As for the pTSFM and FLDM,
the GDR width is calculated by using Eqs. (\ref{pTSFM}) and 
(\ref{GFLDM}), respectively.

\begin{figure}
     \includegraphics[width=9.4cm]{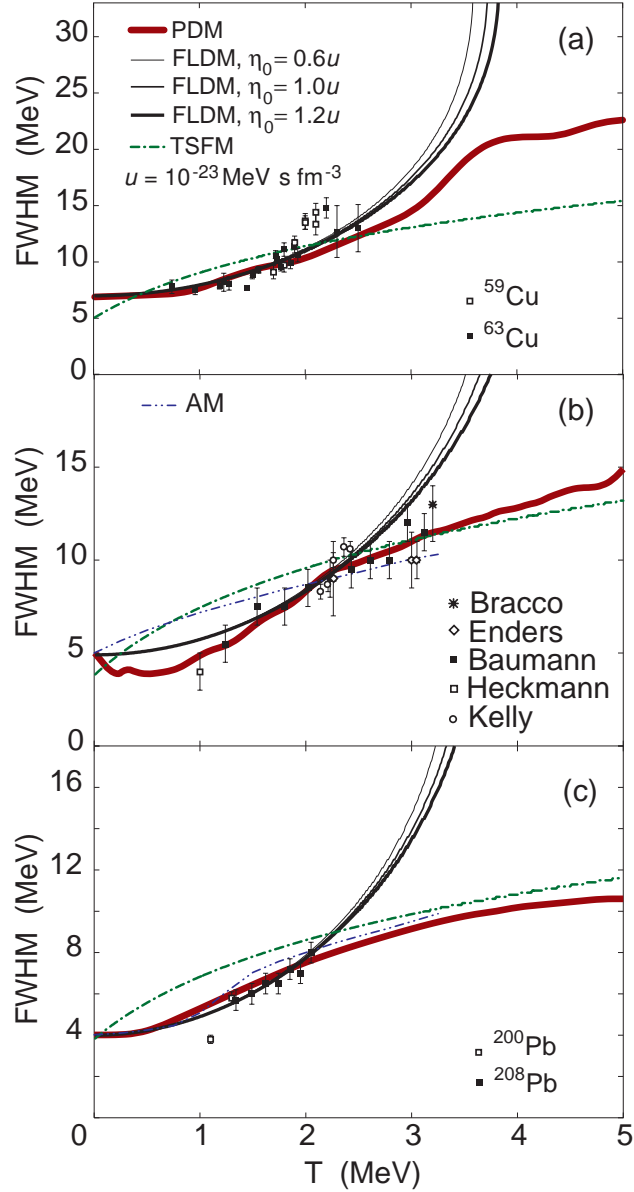}
     \caption{(Color online) FWHM of GDR as functions of $T$  
     for $^{63}$Cu (a), $^{120}$Sn (b), and $^{208}$Pb (c) in 
     comparison with the experimental systematics for 
     for copper (Cu$^{59}$~\cite{Cu59} and 
     Cu$^{63}$~\cite{Cu63}), tin (by Bracco {\it et 
     al.}~\cite{Bracco}, Enders {\it et al.}~\cite{Enders}, 
     Baumann {\it et al.}~\cite{Baumann}, Heckmann {\it et al.}~\cite{Heckmann}, and 
     Kelly {\it et al.}~\cite{Kelly}), and lead 
	 (Pb$^{208}$~\cite{Baumann} and 
	 Pb$^{200}$~\cite{Pb200}) regions. The notations for the theoretical 
     curves are given in (a) and (b). 
     \label{width}}
\end{figure}
Shown in Fig. \ref{width} are the GDR widths predicted by the PDM, 
AM, pTSFM, and FLDM as functions of temperature $T$ 
in comparison with the experimental 
systematics~\cite{Bracco,Enders,Baumann, 
Heckmann,Kelly,Pb200,Cu59,Cu63}, which are also collected 
in Ref. \cite{Schiller}. The PDM predictions fit 
best the experimental systematics for all three nuclei $^{63}$Cu, 
$^{120}$Sn, and $^{208}$Pb. The AM fails to describe the GDR width at 
low $T$ for $^{120}$Sn because thermal pairing was not included in the AM 
calculations, while it slightly overestimates the width for 
$^{208}$Pb (The AM prediction for GDR width in $^{63}$Cu is not 
available). The predictions by the pTSFM is qualitatively similar to 
those by the AM, although to achieve this agreement the pTSFM needs 
to use $\Gamma(0)=$ 5 MeV for $^{63}$Cu and 3.8 MeV for $^{120}$Sn, 
i.e. substantially smaller than the experimental values of around 7 
and 4.9 MeV for $^{63}$Cu and $^{120}$Sn, respectively.  
This model also produces the width saturation similar to 
that predicted by the PDM, although for $^{63}$Cu the width obtained 
within the pTSFM at $T>$ 3 MeV is noticeably smaller than that 
predicted by the PDM. The widths obtained within the FLDM fit the 
data fairy well up to $T\simeq$ 2.5 MeV. However, they do not saturate at 
high $T$, but increases sharply with $T$, and break down at $T_{c}<$ 
4 MeV. As has been mentioned previously in Sec. \ref{FLDM}, 
at $T>$ 2.5 MeV the dependence on $\eta(0)$ (ultimately 
$\alpha$) starts to show up in the FLDM results for the GDR widths, 
which are 18.3, 17.5, and 17 MeV for $\eta_{0}\equiv\eta(0)=$ 0.6, 1.0, and 1.2 $u$, 
respectively, for $^{63}$Cu at $T=$ 3 MeV. The corresponding 
differences between the 
widths obtained by using these values of $\eta(0)$ for 
$^{120}$Sn and $^{208}$Pb are slightly smaller. The values of the 
critical temperature $T_{c}$, starting from which the FLDM width 
becomes imaginary, are 3.58, 3.72, 3.83 MeV by using $\eta(0)=$ 0.6, 1.0, 
and 1.2 $u$, respectively, for $^{63}$Cu. For $^{120}$Sn these 
corresponding values 
for $T_{c}$ are 
3.77, 3.94, and 4.1 MeV, whereas for $^{208}$Pb
they are 3.42, 3.54, and 3.65 MeV, respectively. At these values of 
$T_{c}$ the ratio $\eta(T_{c})/\eta(0)$ is smaller than 3.5, which is not much different 
from the estimation (\ref{ratio2}) (See later in Sec. \ref{eta/s}).
\subsection{Entropy}
\begin{figure}
     \includegraphics[width=9cm]{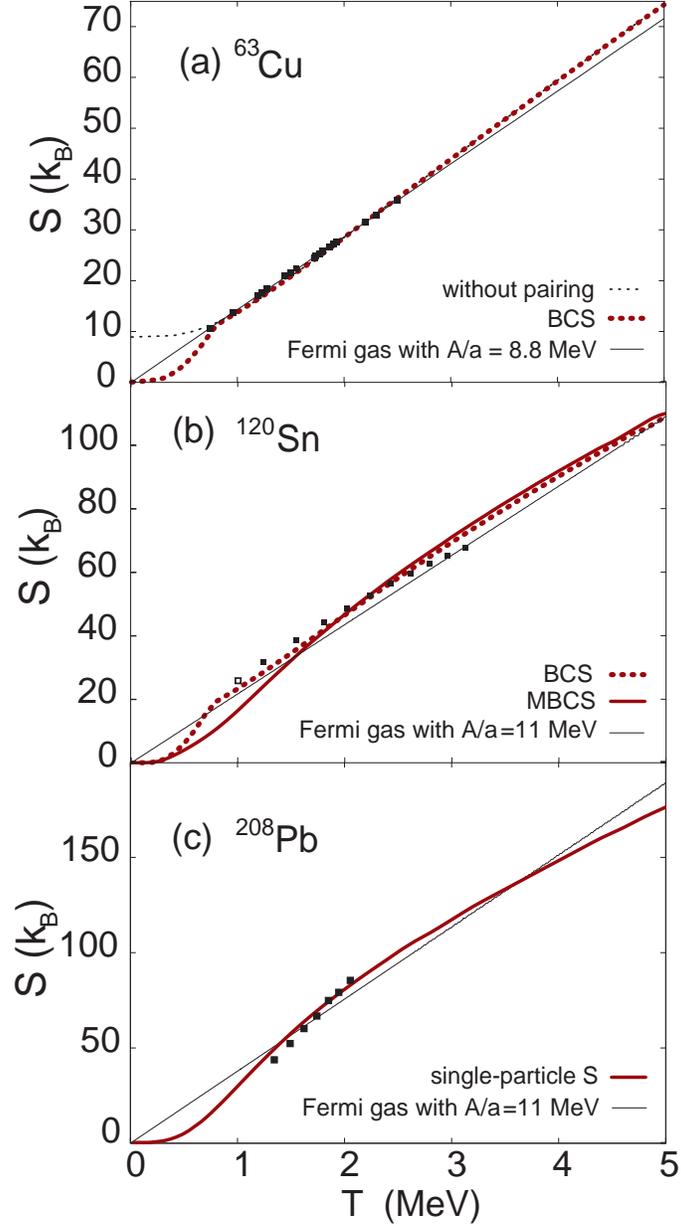}
     \caption{(Color online) Entropies as functions of $T$  
     for $^{63}$Cu (a), $^{120}$Sn (b), and $^{208}$Pb (c) in 
     comparison with the empirical values. The notations are the same 
     as in Fig. \ref{width}.} 
     \label{entropy}
\end{figure}
Compared in Fig. \ref{entropy} are 
the entropies obtained by using the microscopic expressions 
(\ref{Stotal}) and (\ref{SPDM}) and the empirical ones 
extracted from the Fermi-gas formula (\ref{SFG}) by using the 
empirical values for the level-density parameter $a$ discussed 
previously in Sec. \ref{entro}. The microscopic entropy includes 
pairing for open shell nuclei. For $^{63}$Cu, although pairing is not 
included in the calculation of the GDR width, the finite-temperature 
BCS pairing with blocking by the odd proton is taken into account for 
the entropy to ensure its vanishing value at low $T$ [Compare the thick 
dotted line obtained including the BCS pairing and the thin dotted 
line obtained without pairing in Fig. 
\ref{entropy} (a)]. For $^{120}$Sn, the MBCS theory~\cite{MBCS} is 
needed to reproduce the GDR 
width depletion at $T\leq$ 1 MeV in this nucleus due to the 
nonvanishing thermal pairing gap above the temperature of the BCS 
superfluid-normal phase transition [Thick solid line in Fig. \ref{width} 
(b)] so the MBCS thermal pairing gap is also included 
in the calculation of the entropy. 
For the closed-shell nucleus $^{208}$Pb, the 
quasiparticle entropy $S_{F}$ in Eq. (\ref{SPDM}) becomes the 
single-particle entropy because of the absence of pairing. The good agreement between the 
results of microscopic calculations  and the empirical extraction 
indicates that the level-density parameter for $^{63}$Cu within the 
temperature interval 0.7 $<T<$ 2.5 MeV can be considered to be 
temperature-independent and equal to $a=63/8.8\simeq$ 7.16 MeV$^{-1}$, 
whereas for $^{120}$Sn and $^{208}$Pb the level-density parameter 
varies significantly with $T$~\cite{Baumann}. The Fermi-gas entropy $S^{\rm FG}$ 
(\ref{SFG}) with a constant level-density parameter $a$ fits best 
the microscopic and empirical results with $A/a=$ 8.8 MeV for 
$^{63}$Cu, and 11 MeV for $^{120}$Sn and $^{208}$Pb. 
\subsection{Ratio $\eta/s$}
\label{eta/s}
\subsubsection{Model-dependent predictions vs empirical results}
\begin{figure}
     \includegraphics[width=16.0cm]{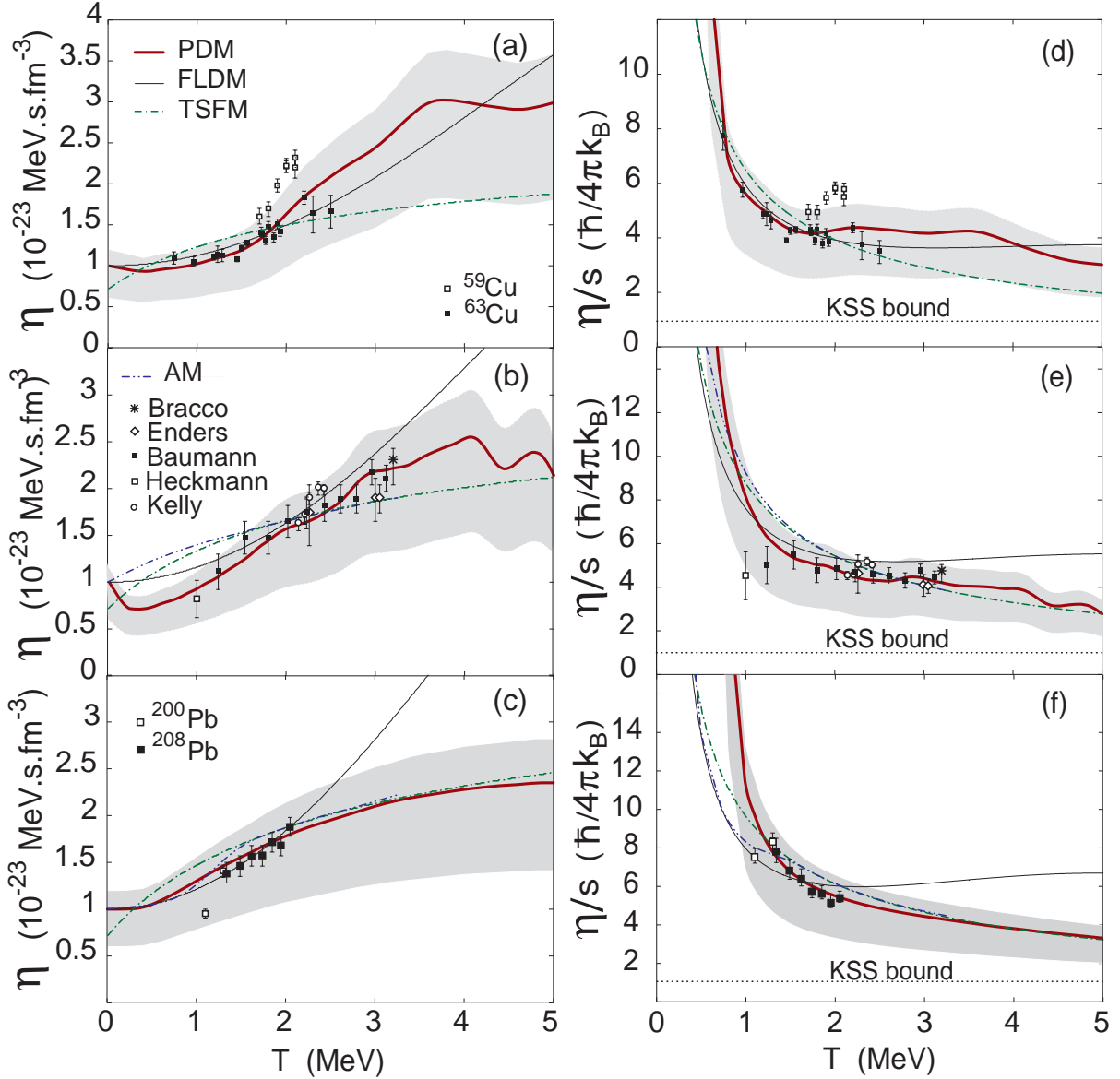}
     \caption{(Color online) Shear viscosity $\eta(T)$ [(a) - (c)] and ratio 
     $\eta/s$ [(d) - (f)] as functions of $T$ for nuclei in copper [(a) and 
     (d)], tin [(b) and (e)], and lead [(c) and (f)] regions. The  
     gray areas are the PDM predictions by using $0.6u\leq\eta(0)\leq 
     1.2u$. The same notations as in Fig. 1 are used to denote 
     the empirical results, which are extracted by using the 
     corresponding experimental widths and energies 
     for the GDR in copper~\cite{Cu59,Cu63}, 
     tin~\cite{Bracco,Enders,Baumann,Heckmann,Kelly}, 
     and lead~\cite{Pb200,Baumann} regions.}
     \label{eta&ratio}
\end{figure}
The predictions for the shear viscosity $\eta$ and the ratio $\eta/s$ by the PDM, FLDM, AM, and 
pTSFM for $^{63}$Cu, $^{120}$Sn, and $^{208}$Pb are plotted as functions of $T$ 
in Fig. \ref{eta&ratio} in comparison with the empirical results. 
The empirical values for $\eta$ in Figs. \ref{eta&ratio} (a) - 
\ref{eta&ratio} (c) are extracted from the 
experimental systematics for GDR in copper, tin and lead regions~
\cite{Bracco,Enders,Baumann,Heckmann,Kelly,Pb200,Cu59,Cu63,Schiller} making 
use of Eq. (\ref{eta1}). 
The PDM predictions for $\eta$ [thick solid lines and gray areas] are 
obtained from Eq.
(\ref{eta1}) by using the temperature-dependent 
GDR widths from in Fig. \ref{width}, and 
$E_{GDR}(T)$, which oscillates slightly around $E_{GDR}(0)$ as
$T$ varies [See Fig. 4 (b) of Ref. \cite{PDM2}].  The predictions by the FLDM and AM 
are obtained by using the same resonance energy 
$\hbar\omega\equiv E_{GDR}=E_{GDR}(0)$ with $\eta(0)$ = 1$u$ and 
$A/a=$ 11 MeV because this value of $A/a$ gives the best fit to
experimentally extracted entropy, as shown in Fig. 
\ref{entropy}~\footnote{In Ref.~\cite{Auerbach} the values 
$\hbar\omega=$ 20 MeV and $\alpha=$ 9.2, which correspond to the isoscalar 
mode, were used for Eqs. 
(\ref{etaFLDM}) -- (\ref{zs}). The present article extracts 
$\eta/s$ from the GDR,  so the use of $\hbar\omega=E_{GDR}(0)$ 
and $\eta(0)=$ 1$u$ is appropriate as it corresponds to $\alpha=$ 
4.27, close to the value 4.6 for the isovector mode within the 
FLDM~\cite{alpha}.}

It is seen in Fig. \ref{eta&ratio} that the predictions by the PDM have the best overall 
agreement with the empirical results for all three nuclei $^{63}$Cu, 
$^{120}$Sn, and $^{208}$Pb. The PDM produces an increase 
of $\eta(T)$ with $T$ up to 3 - 3.5 MeV and a saturation of $\eta(T)$ within (2 - 3)$u$ 
at higher $T$ [with 
$\eta(0)=$ 1$u$]. The ratio $\eta/s$ decreases sharply with 
increasing $T$ up to $T\sim$ 1.5 MeV, starting from which the decrease 
gradually slows down to reach (2 - 3) KSS units 
at $T=$ 5 MeV. The FLDM has a similar trend as that of the PDM up to 
$T\sim$ 2 - 3 MeV, but at higher $T$ ($T>$ 3 MeV for $^{120}$Sn or 2 MeV for 
$^{208}$Pb) it produces an increase of both $\eta$ and $\eta/s$ with $T$. 
At $T=$ 5 MeV the FLDM model predicts the ratio $\eta/s$ within (3.7 - 6.5) KSS units, which are 
roughly 1.5 times - twice larger than the PDM predictions.

The AM and pTSFM show a similar trend for $\eta$ and $\eta/s$. 
However, in order to obtain such similarity, $\eta(0)$ in the pTSFM 
calculations has to be reduced to 0.72$u$ instead of 1$u$. They all 
overestimate $\eta$ at $T<$ 1.5 MeV. Because of the smaller $\eta(0)$ 
in use, the pTSFM predicts a much lower saturated value for $\eta$ at high 
$T$ for the lighter nucleus $^{63}$Cu, and  consequently, a smaller
$\eta/s$, which amounts to around 2$u$ at $T=$ 5 MeV, i.e. comparable 
to the PDM's prediction by using $\eta(0)=$ 0.6$u$.

\begin{figure}
     \includegraphics[width=15.0cm]{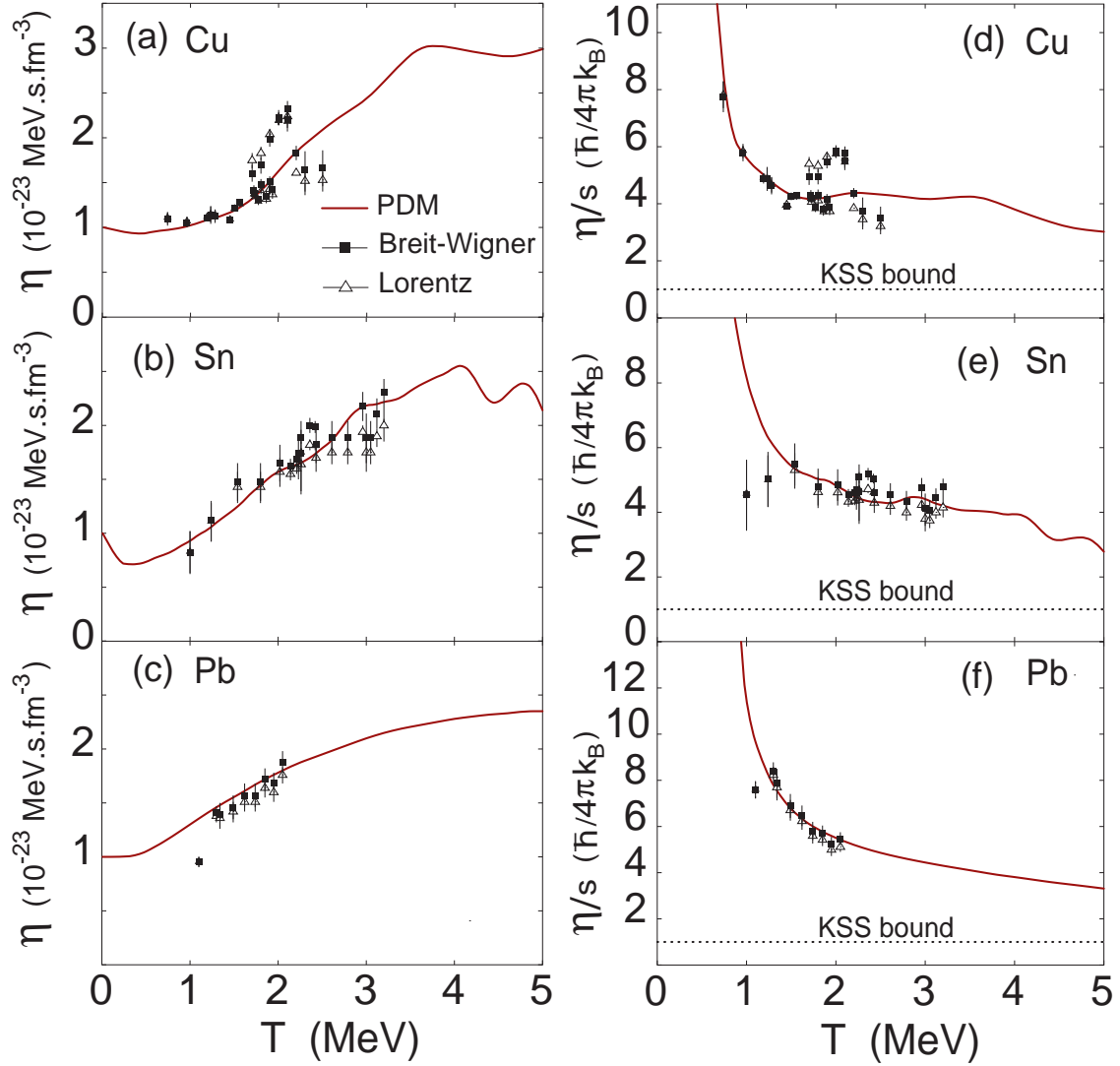}
     \caption{(Color online) Shear viscosity $\eta(T)$ and ratio 
     $\eta/s$ as functions of $T$ for nuclei in copper [(a) and 
     (d)], tin [(b) and (e)], and lead [(c) and f)] regions. The 
     solid boxes and open triangles with error bars 
     denote the empirical results obtained by 
     using Eqs. (\ref{eta1}) and 
     (\ref{etaLor}), respectively. The solid lines are the PDM 
     predictions for $^{63}$Cu, $^{120}$Sn, and $^{208}$Pb, as in Fig. \ref{eta&ratio}.}
     \label{Lor&BW}
\end{figure}
The use of the Lorentz distribution instead of the Breit-Wigner one 
for the photoabsorption cross section does not cause a significant 
difference for $\eta$ and hence, $\eta/s$. As shown in Fig. 
\ref{Lor&BW}, using Eq. (\ref{etaLor}) instead of Eq. (\ref{eta1})
leads to some slight increase of $\eta$ and $\eta/s$ at low $T$ and 
decrease of them at high $T$. Depending on the competition between 
$\Gamma(T)/\Gamma(0)$, which increases with $T$, and 
$\{E_{GDR}(0)^{2}-[\Gamma(0)/2]^{2}+[\Gamma(T)/2]^{2}\}^{-2}$, which 
decreases as $T$ increases, the values of $\eta$ and $\eta/s$ 
obtained by using Eq. (\ref{etaLor}) can be larger or smaller
than those predicted by Eq. (\ref{eta1}). For example, Eq. (\ref{etaLor}) 
leads to slightly larger $\eta$ and $\eta/s$ for $^{59}$Cu at $T<$ 2 
MeV, but smaller values for these quantities for $^{63}$Cu at $T>$ 2 
MeV. For $^{120}$Sn and $^{208}$Pb, the Lorentz distribution of the 
photobasroption cross section produces slightly smaller $\eta$ and 
$\eta/s$ at high $T$.
\subsubsection{Model-independent assessment}
A model-independent estimation for the high-$T$ limit of the ratio $\eta/s$ can be inferred 
directly from Eqs. (\ref{eta1}) and (\ref{SPDM}) under the assumption of GDR 
width saturation as follows. From the trend of the GDR width's increase, predicted by the PDM, AM and 
pTSFM shown in Fig. \ref{width}, it can be assumed that
at the highest $T_{max}\simeq$ 5 - 6 MeV where the GDR can still exist, the 
GDR width $\Gamma(T)$ cannot exceed $\Gamma_{max}\simeq 
3\Gamma(0)\simeq 0.9E_{GDR}(0)$~\cite{Auerbach1}. 
Because the GDR energy $E_{GDR}(T)$ is stable against the variation of 
$T$, one can also put $E_{GDR}(T)\simeq E_{GDR}(0)$. Inserting these 
values into Eq. (\ref{eta1}), the high-$T$ limit of $\eta(T)$ is 
found as
\begin{equation}
    \eta_{max} \simeq 2.551\times\eta(0)~.
    \label{etamax}
    \end{equation}
The high-$T$ limit of the entropy density $s$ is obtained by noticing 
that, $S_{F}\to 2\Omega\ln{2}$ at $T\to\infty$ 
because $n_{j}\to$ 1/2, where $\Omega=\sum_{j}(j+1/2)$ for 
the spherical single-particle basis or sum of all doubly-degenerate levels for the 
deformed basis. The particle-number conservation requires that $A = 
\Omega$ since all single-particle occupation numbers are equal to 
1/2. This leads to the following high-$T$ limit of entropy density 
$s$ (\ref{dens}):
\begin{equation}
    s_{max} = 2\rho\ln{2}\simeq 0.222~(k_{B})~.
    \label{smax}
    \end{equation} 
    Dividing the right-hand side of Eq. (\ref{etamax}) by that of Eq. 
(\ref{smax}) yields the high-$T$ limit (or lowest bound) for $\eta/s$ 
in finite nuclei
\begin{equation}
    \bigg(\frac{\eta}{s}\bigg)_{min}\simeq 2.2^{+0.4}_{-0.9}~({\rm 
    KSS}~{\rm units})~,
\label{R1}    
\end{equation}
by using the empirical values for $\eta(0) = 
1.0^{+0.2}_{-0.4}~u$~\cite{Auerbach1,fission}.

To ensure the validity of this result, an alternative estimation is carried out 
by using the Fermi-gas entropy without assuming a width 
saturation~\cite{Kelly}. From Eq. (\ref{eta1}) it follows 
that the ratio $\eta(T)/s$ is not smaller than the KSS bound (\ref{KSSbound})
only if the GDR width $\Gamma(T)$ takes the
values between $\Gamma_{1}\leq\Gamma\leq\Gamma_{2}$, where
\begin{equation}
\Gamma_{1,2}=\eta(0)\frac{4E_{GDR}(0)^{2}+\Gamma(0)^{2}}{2sK\Gamma(0)}    
\bigg\{1\pm\sqrt{1-\bigg[\frac{4sK\Gamma(0)E_{GDR}(T)}
{\eta(0)(4E_{GDR}(0)^{2}+\Gamma(0)^{2})}\bigg]^{2}}\bigg\}~,\hspace{2mm} 
K=\frac{\hbar}{4\pi k_{B}}~.
    \label{Gamma12}
    \end{equation}
Inserting into Eq. (\ref{Gamma12}) the entropy density (\ref{dens}) 
with $S$ given by the Fermi-gas formula (\ref{SFG}), where 
the lower and upper bounds for $A/a$  are taken equal to 8 and
13 (MeV), respectively, for finite hot nuclei~\cite{MBCS,Hagen}, one 
finds for $\eta(0)=1.0 u$ that, at $T=$ 6 MeV, 
the GDR width $\Gamma$ should be confined
within the intervals 1.3 $\leq\Gamma/\Gamma_{0}\leq$ 34.8
for $A/a=$ 8 MeV, and
0.8 $\leq\Gamma/\Gamma_{0}<$ 58 for $A/a=$ 13 MeV. The 
values for $\eta/s$ found at the middle of these intervals with
$\Gamma/\Gamma_{0}\simeq$ 18 and 29.4 amount to
${\eta}/{s}\simeq$ 1.76 KSS units for $A/a=$ 8 MeV, and 1.9 KSS units for
$A/a = $ 14 MeV. 
Including the error bars produced by the lower and upper values of 
$\eta(0)$ above, one finds
\begin{equation}
    \frac{\eta}{s}\simeq 1.79^{+0.04}_{-0.34}~~{\rm 
    and}~~1.9^{+0.03}_{-0.15}~({\rm 
        KSS}~{\rm units})~,
    \label{R2}
    \end{equation}   
for $A/a = $ 8 and 14 MeV, respectively. The limit (\ref{R2}) 
turn out to be within the high-$T$ limit (\ref{R1}).
        
\section{Conclusions}
In the present article, by using the Kubo relation and the 
fluctuation-dissipation theorem, the shear viscosity $\eta$ and the ratio $\eta/s$ have been extracted from the
experimental systematics for the GDR widths in copper, tin and lead regions at
$T\neq$ 0, and compared with the theoretical predictions by four 
independent theoretical models. The calculations adopt the value 
$\eta(0)=1.0_{-0.4}^{+0.2}\times u$ ($u = 10^{-23}$ Mev s fm$^{-3}$) as a parameter,  
which has been extracted by fitting the giant resonances at $T=$ 0~\cite{Auerbach1} and 
fission data~\cite{fission}. The analysis of numerical calculations show that 
the shear viscosity $\eta$ increases between (0.5 - 2.5)$u$ with 
increasing $T$ from 0.5 up to $T\simeq$ 3 - 3.5 
MeV for $\eta(0)=$ 1$u$. At higher $T$ the PDM, AM, and pTSFM predict a saturation, or at 
least a very slow increase of $\eta$, whereas the FLDM show 
a continuously strong increase of $\eta$, with $T$. At $T=$ 5 MeV, 
the PDM estimates $\eta$ between around (1.3 - 3.5)$u$.

All theoretical models predict a decrease of the 
ratio $\eta/s$ with increasing $T$ up to $T\simeq$ 2.5 MeV. At higher 
$T$, the PDM, AM, and pTSFM show a continuous decrease of $\eta/s$, whereas the 
FLDM predicts an increase of $\eta/s$, with increasing $T$.
The PDM fits best the empirical values for $\eta/s$ extracted at 
0.7$\leq T\leq$ 3.2 MeV for all three nuclei, $^{63}$Cu $^{120}$Sn, 
and $^{208}$Pb. At $T=$ 5 MeV, the values of $\eta/s$ predicted by 
the PDM reach $3^{+0.63}_{-1.2}$, $2.8^{+0.5}_{-1.1}$, 
$3.3^{+0.7}_{-1.3}$ KSS units for $^{63}$Cu, $^{120}$Sn, and 
$^{208}$Pb, respectively. Combining these results with the 
model-independent estimation for the high-$T$ limit of $\eta/s$, 
which is $2.2^{+0.4}_{-0.9}$ KSS units, one can conclude that the 
value of $\eta/s$ for medium and heavy nuclei at $T=$ 5 MeV is in 
between (1.3 - 4.0) KSS units, which is about (3 - 5) times smaller 
(and of much less uncertainty) that the value between (4 - 19) KSS units predicted by 
the FLDM for heavy nuclei, where the same lower value $\eta(0)=$ 
0.6$u$ was used. By using the same upper value $\eta(0)=$ 2.5$u$ as in 
Ref. \cite{Auerbach}, instead of $\eta(0)=$ 1.2$u$, this interval for
$\eta/s$ becomes (1.3 - 8.3) KSS units, whose uncertainty of 7 KSS 
units is still smaller than that predicted by the FLDM (15 KSS units). 
This estimation also indicates that nucleons inside a hot 
nucleus at $T=$ 5 MeV has nearly the same ratio $\eta/s$ as that 
of QGP, around (2 - 3) KSS units, at $T>$ 170 MeV discovered at RHIC and LHC.

\acknowledgments
The numerical calculations were carried out using the FORTRAN IMSL
Library by Visual Numerics
on the RIKEN Integrated Cluster of Clusters (RICC) system. Thanks are 
due to G.F. Bertsch and P. Danielewicz for stimulating discussions, 
as well as N. Quang Hung for assistance in numerical calculations. 

\end{document}